\documentclass[osajnl,twocolumn,showpacs,superscriptaddress,10pt]{revtex4-1} 
\usepackage{amsmath,amssymb,graphicx}

\begin{document}

\title{Long Range Surface Plasmons in Multilayer Structures}
\author{Aida Delfan}\email{Corresponding author:adelfan@physics.utoronto.ca}
\author{J. E. Sipe}
\affiliation{Department of Physics and Institute for Optical Sciences, University of Toronto, 60 St. George Street, Ontario M5S 1A7,Canada}

\begin{abstract}
We present a new strategy, based on a Fresnel coefficient pole analysis, for designing an asymmetric multilayer
structure that supports long range surface plasmons (LRSP). We find that
the electric field intensity in the metal layer of a multilayer LRSP structure
can be even slightly smaller than in the metal layer of the corresponding
symmetric LRSP structure, minimizing absorption losses and resulting in
LRSP propagation lengths up to $2mm$. With a view towards biosensing
applications, we also present semi-analytic expressions for a standard
surface sensing parameter in arbitrary planar resonant structures, and in
particular show that for an asymmetric structure consisting of a gold film
deposited on a multilayer of SiO$_{2}$ and TiO$_{2}$ a
surface sensing parameter $G=1.28nm^{-1}$ can be achieved.
\end{abstract}


\maketitle

\section{Introduction}

Surface plasmon polaritons (or ``surface plasmons'' (SP) for short) are
excitations at metal-dielectric interfaces involving both electronic and
electromagnetic degrees of freedom \cite{SP}. \ They have attracted interest
for a wide range of applications, including waveguiding and biosensing \cite%
{Barnes,SP_sensing_rev,berini_sensing}, despite the fact that absorption in
the metal typically restricts their propagation length to only a few tens of
microns. \ These losses can be minimized if a thin metal layer is situated
between two media -- a cladding above and a substrate below -- with the same
dielectric constant. In this symmetric structure the SPs at the two
metal-dielectric interfaces can couple and form an excitation in which most
of the energy is in the bounding dielectrics, resulting in small propagation
losses and propagation lengths of the order of a few millimeters. \ For this
reason, these excitations are called long range surface plasmons (LRSP)\cite%
{LRSP,stegeman}. \ They can survive if the cladding and substrate dielectric
constants are not identical, as long as the difference is not too great. \ 

When the cladding above the metal film is a gas or liquid, as in biosensing
applications, it is challenging to find a solid substrate that comes close
to matching the cladding dielectric constant. \ If the cladding is an
aqueous solution, Teflon and Cytop are two of the few candidate materials
that can provide such a match \cite{cytop,teflon}. \ An alternate approach
is to design an asymmetric layered structure that can support LRSP; examples
include a suspended waveguide structure \cite{Berini1} and 1D photonic
crystal structures \cite{Konopsky1,Konopsky2}. In this paper, we present a new
approach to the design of asymmetric structures for LRSPs, based on a
Fresnel coefficient pole analysis. Motivated by biosensing applications, we
consider a water cladding, and focus on minimizing the electric field in the
metal in order to reduce the losses. \ In a scenario where the structure is
used for sensing the presence of a layer of molecules adsorbed onto the
metal film from the water, we calculate the value of a standard surface
sensing parameter \cite{berini_sensing, Berini1} that results, and show that
in the limit of thin molecular layers it can be derived from a semi-analytic
expression that also follows from the pole analysis.

The organization of this paper is as follows: In Sec. \ref{design} we present our
approach for designing a periodic multilayer structure supporting LRSPs. In
Sec. \ref{LRSP_fields} we compare the field intensity profile of the LRSP supported by the
multilayer structure to the LRSP in a symmetric structure, and calculate
absorption and coupling losses for finite symmetric and asymmetric
structures. In Sec. \ref{sensing} we derive a semi-analytic expression for a standard
sensing parameter used to characterize the effect of thin molecular layers
on the optical properties of arbitrary planar resonant structures, and
compare exact and approximate calculations for the structures studied in
Sec. \ref{LRSP_fields}. The paper ends with our conclusions and a comparison with related
work.

\section{Asymmetric multilayer structures for LRSPs}\label{design}

\begin{figure}[htbp]
\centerline{\includegraphics[width=7.0cm]{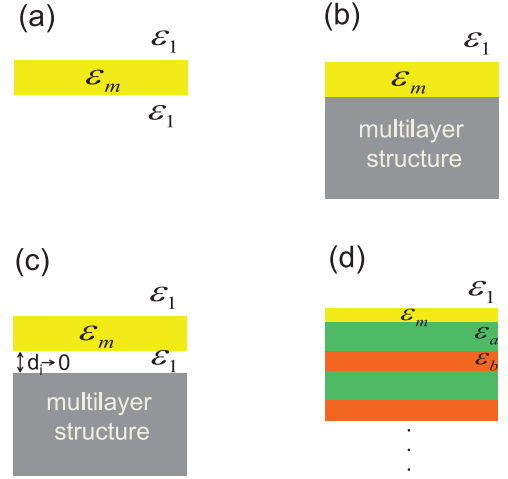}}
\caption{(a) A symmetric structure for LRSP, with a metal layer between two
media with the same dielectric constant, (b) An asymmetric structure with a multilayer substrate beneath the metal layer, (c) an infinitesimally thin layer of medium with dielectric constant $\varepsilon_{1}$ inserted under the metal layer, and (d) an asymmetric structure with a metal film supported by an infinite periodic structure consisting of layers of dielectric constant $\varepsilon_{a}$ and $\varepsilon_{b}$.}
\label{Fig:design}
\end{figure}

As a reference we begin by considering a symmetric structure, consisting of
a thin layer of metal of thickness $d_{m}$ with a dielectric constant $%
\varepsilon _{m}$, sandwiched between a cladding and substrate of the same
dielectric constant $\varepsilon _{1}$ (Fig. \ref{Fig:design}a). The
Fresnel reflection coefficient for light incident on the metal from the
cladding, with wavevector component $\kappa $ parallel to the interfaces, is 
\begin{equation}\label{FC_ref_sym}
R_{cs}=r_{1m}+\frac{t_{1m}r_{m1}t_{m1}e^{2iw_{m}d_{m}}}{%
1-r_{m1}r_{m1}e^{2iw_{m}d_{m}}}, 
\end{equation}%
where $w_{i}=\sqrt{{\tilde{\omega}}^{2}\varepsilon _{i}-\kappa ^{2}}$ and $%
\tilde{\omega}=\omega /c$ \cite{formalism1}. For real $\kappa $ we
define the square root according to $\text{Im}\sqrt{Z}\geq 0$, with $\text{Re
}\sqrt{Z}\geq 0$ if $\text{Im}\sqrt{Z}=0$; this guarantees that in the limit 
$z\rightarrow \pm \infty $ the reflected and transmitted fields, following
respectively from Eq. \eqref{FC_ref_sym} and the corresponding transmission
coefficient, are either evanescent moving away from the structure, or carry
energy away from it. \ The $r_{ij}$ and $t_{ij}$ are respectively the
Fresnel reflection and transmission coefficients from medium $i$, with
dielectric constant $\varepsilon _{i}$, to medium $j$, with dielectric
constant $\varepsilon _{j}$. \ For $s$-polarized light the coefficients are%
\begin{eqnarray}\label{spol}
r_{ij} &=&\frac{w_{i}-w_{j}}{w_{i}+w_{j}},   \\
t_{ij} &=&\frac{2w_{i}}{w_{i}+w_{j}},  \notag
\end{eqnarray}%
and for $p$-polarized light they are 
\begin{eqnarray}\label{ppol}
r_{ij} &=&\frac{w_{i}\varepsilon _{j}-w_{j}\varepsilon _{i}}{%
w_{i}\varepsilon _{j}+w_{j}\varepsilon _{i}},   \\
t_{ij} &=&\frac{2n_{i}n_{j}w_{i}}{w_{i}\varepsilon _{j}+w_{j}\varepsilon _{i}%
},  \notag
\end{eqnarray}%
where $n_{i}\equiv \sqrt{\varepsilon _{i}}$.

Surface electromagnetic resonances are generally signalled by poles in
the Fresnel coefficients, indicating that fields can exist near the surface
in the absence of incident light. \ Thus the condition for the LRSP
excitation is 
\begin{equation}
1-r_{m1}r_{m1}e^{2iw_{m}d_{m}}=0.  \label{pole1}
\end{equation}%
For $p$-polarized light this can be satisfied at a resonance wavenumber $%
\kappa _{res}^{sym}$, which is complex due to absorption in the metal. \ The
real part of $\kappa _{res}^{sym}$ is greater than $\tilde{\omega}n_{1}$,
indicating a field structure bound to the region of the thin film. \ We note
that the extension of $\kappa $ from the real axis to the complex plane can
introduce subtleties associated with the definition of the square root in $%
w_{i}(\kappa )$; we will turn to those in Sec. \ref{LRSP_fields}, but they will not
affect the discussion in this section. \ 

As an example of a symmetric structure we consider a gold metal layer with a
thickness $d_{m}=20nm$, with water as the cladding and substrate. \ \ At a
wavelength of $\lambda =1310nm$, and with dielectric constants of water and
gold taken as $\varepsilon _{water}=(1.3159+i1.639\times 10^{-5})^{2}$ \cite%
{Berini1} and $\varepsilon _{gold}=-86.08+i8.322$ \cite{Berini1}, a
numerical search of $\kappa $ in the complex plane identifies the LRSP by
finding the complex value $\kappa _{res}^{sym}$ where (\ref{pole1}) is
satisfied for $p$-polarized light; we find an effective index for the LRSP
of $n_{eff}^{sym}=1.31829+i5.34\times 10^{-5}$, where $n_{eff}^{sym}=\kappa
_{res}^{sym}/\tilde{\omega}$. \ The very small imaginary part leads to a
mode loss of about 2.23dB/mm (also considered by Min et al. \cite{Berini1}), or
equivalently an energy propagation length of about $1.95mm$.

For any typical LRSP symmetric structure, such as the one above, we suppose
that $\kappa _{res}^{sym}$ has been found. \ We now want to design an
asymmetric structure (Fig. \ref{Fig:design}b), where the substrate has been
replaced by a multilayer, to support LRSPs that mimic those of Fig. \ref{Fig:design}a. \ 
Any surface electromagnetic resonances in this new structure are signalled
by poles in the Fresnel coefficients, for example in the reflection
coefficient for light incident from the cladding, $\bar{R}_{1s}$, 
\begin{equation}
\bar{R}_{1s}=r_{1m}+\frac{t_{1m}\bar{R}_{ms}t_{m1}e^{2iw_{m}d_{m}}}{1-r_{m1}%
\bar{R}_{ms}e^{2iw_{m}d_{m}}},  \label{reflection}
\end{equation}%
where $\bar{R}_{ms}$ is the Fresnel reflection coefficient for light
incident from a semi-infinite metal placed above the multilayer structure of
interest in Fig. \ref{Fig:design}b. The reflection coefficient, $\bar{R}_{1s}$, has a pole
when 
\begin{equation}
1-r_{m1}\bar{R}_{ms}e^{2iw_{m}d_{m}}=0.  \label{pole2}
\end{equation}%
This will clearly lead to a $\kappa _{res}$ equal to $\kappa _{res}^{sym}$
(recall Eq. \eqref{pole1}), if 
\begin{equation}
\bar{R}_{ms}=r_{m1},  \label{eqivcondition}
\end{equation}%
where the Fresnel coefficients are evaluated at $\kappa _{res}^{sym}$; we
take this as our design target. \ We can simplify it so that it only
involves the properties of the metal layer through a dependence on $\kappa
_{res}^{sym}$ by the following strategy: \ Insert an infinitesimally thin
layer, with a thickness of $d_{i}\rightarrow 0$ and dielectric constant $%
\varepsilon _{1}$, between the metal layer and the multilayer (Fig. \ref%
{Fig:design}c). \ Then $\bar{R}_{ms}$ is easily found to be 
\begin{equation}
\bar{R}_{ms}=r_{m1}+\frac{t_{m1}\mathcal{R}_{1s}t_{1m}}{1-r_{1m}\mathcal{R}%
_{1s}},  \label{reflection}
\end{equation}%
where $\mathcal{R}_{1s}$ is the Fresnel reflection coefficient for light
incident from the cladding to the multilayer structure, when there is no
metal layer present. From Eq. %
\eqref{reflection}, it is clear that \eqref{eqivcondition} is satisfied if $%
\mathcal{R}_{1s}=0$ at $\kappa _{res}^{sym}$. \ 

We now specialize to a periodic multilayer structure consisting of layers $a$
and $b,$ with respectively thicknesses $d_{a}$ and $d_{b}$ and real
dielectric constants $\varepsilon _{a}$ and $\varepsilon _{b}$. For an
infinite periodic structure, the Fresnel coefficient $\mathcal{R}_{1s}$ is
given by 
\begin{eqnarray}
\mathcal{R}_{1s} &=&r_{1a}+\frac{t_{1a}R_{per}t_{a1}}{1-r_{a1}R_{per}}, 
\notag  \label{reflection1} \\
&=&\frac{R_{per}-r_{a1}}{1-r_{a1}R_{per}},
\end{eqnarray}%
where we have assumed the top layer is of type $a$, and $R_{per}$ is the
reflection coefficient for light incident on the periodic structure from a
semi-infinite medium of dielectric constant $\varepsilon _{a}$ (see Fig. \ref{Fig:design}d). In the second line of Eq. 
\eqref{reflection1}, we have used the Fresnel coefficient identities $%
t_{ij}t_{ji}-r_{ij}r_{ji}=1$ and $r_{ij}=-r_{ji}$ \cite{formalism2}, and from that line we find that the
condition $\mathcal{R}_{1s}=0$ is satisfied if 
\begin{equation}
R_{per}=r_{a1},  \label{ra1}
\end{equation}%
which we refer to as our \textit{matching condition}. \ When considering the
propagation of the LRSP, if the multilayer is to simulate a uniform
substrate with dielectric constant equal to that of the cladding, or at
least nearly so, this condition must be satisfied, or nearly satisfied, when
the Fresnel coefficients are evaluated at $\kappa _{res}^{sym}$. \ 

To establish a protocol for designing such a multilayer, it is useful to
begin by neglecting all loss in the cladding and the metal, and any that
might be present in the multilayer. \ In this approximation $\varepsilon _{1}
$ is replaced by its real part, and $\kappa _{res}^{sym}$ and $n_{eff}^{sym}$
are replaced by their real parts.\ This lossless approximation will allow us
to winnow down the parameter space easily, to the point that the design can
be completed; \textit{thus until the last three paragraphs of this section
we assume }$\kappa _{res}^{sym}=\tilde{\omega}n_{eff}^{sym}$ \textit{and }$%
\varepsilon _{1}$ \textit{to be real in our analyses, and use the real parts
of the actual quantities in our calculations.}

Since $n_{eff}^{sym}>n_{1}$ we immediately have $\left\vert
r_{a1}\right\vert =1$, indicating that the field is evanescent in the
cladding, and thus from \eqref{ra1} we must have $\left\vert
R_{per}\right\vert =1$. \ If the dielectric constants of the layer materials
are large enough so that the fields are propagating within the layers
themselves ($w_{a}$ and $w_{b}$ real), this requires that at $\kappa
_{res}^{sym}$ we are within one of the photonic band gaps of the multilayer
structure, so that the overall field structure is evanescent in the
multilayer, and the reflectivity $R_{per}$ is of unit norm. \ To identify
the condition for this to be so, note that the unit cell transfer matrix of
the periodic multilayer structure is $%
m_{unit}=m_{a}(d_{a})m_{ab}m_{b}(d_{b})m_{ba}$, with $m_{ij}$ the interface
transfer matrix between medium $i$ and $j$, and $m_{i}(d_{i})$ the
propagation transfer matrix in medium $i$ \cite{formalism2}; $R_{per.}$ is calculated from the eigenvector of $m_{unit}$, 
\begin{equation}
m_{unit}\left( 
\begin{array}{c}
R_{per.} \\ 
1 \\ 
\end{array}%
\right) =\lambda \left( 
\begin{array}{c}
R_{per.} \\ 
1 \\ 
\end{array}%
\right) ,
\end{equation}%
with eigenvalue $\lambda =e^{i\mu L}$, where $L=d_{a}+d_{b}$, and $\mu $ is
the complex Bloch wavenumber. For an overall field structure that is
evanescent in the multilayer structure, signalling that we are in a photonic
band gap, we must have $|\lambda |>1$ \cite{yariv}. Writing the unit cell
matrix elements by $A$, $B$, $C$, and $D$, 
\begin{equation}
m_{unit}=\left( 
\begin{array}{cc}
A & B \\ 
C & D 
\end{array}%
\right) ,  \label{munit}
\end{equation}%
we have 
\begin{equation}
R_{per}=\frac{B}{\lambda -A},
\end{equation}%
with 
\begin{equation}
\lambda =\frac{A+D}{2}\pm \sqrt{\big(\frac{A+D}{2}\big)^{2}-1}.
\end{equation}%
\cite{yariv}. The matrix elements of \eqref{munit} are found by multiplying
the transfer matrices of which it is composed; in particular we find 
\begin{eqnarray}
\frac{A+D}{2}&=&\frac{1}{1-r_{ba}^{2}}\big(\cos(w_{a}d_{a}+w_{b}d_{b}) \notag \\
&&-r_{ba}^{2}\cos (w_{a}d_{a}-w_{b}d_{b}).\big)
\label{AplusD}
\end{eqnarray}%
Location within a band gap ($\mu $ purely imaginary, $\lambda $ real) is
thus signalled by $|A+D|/2>1$. At the band edges, $\lambda =\pm 1$ and $%
|A+D|/2=1$ \cite{yariv}. \ It is within the band gaps that we have $\left\vert R_{per.}\right\vert =1$,
and it is there we must seek to satisfy Eq. \eqref{ra1}.

\begin{figure}[htbp]
\centerline{\includegraphics[width=7.5cm]{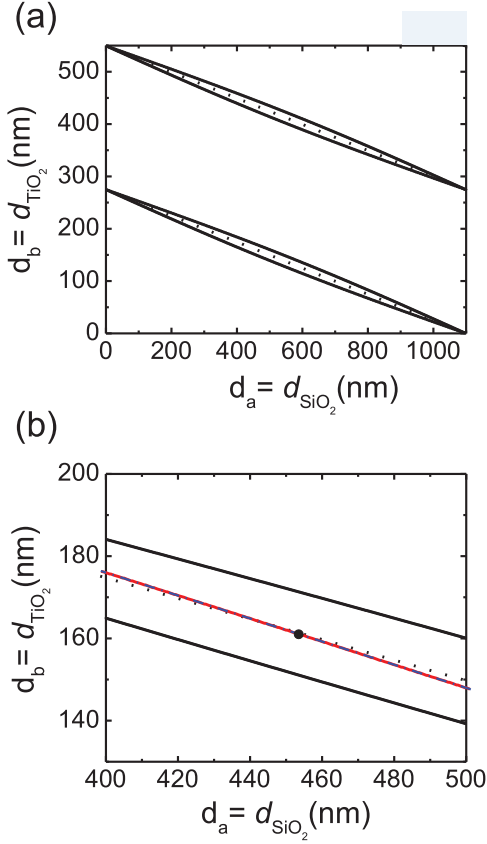}}
\caption{ (a) The $d_{SiO_{2}}$ and $d_{TiO_{2}}$ (solid black lines) for which $\kappa %
_{res}^{sym}$ is on a bandedge of the periodic multilayer structure in the lossless case , and the straight lines of $w_{SiO_{2}}d_{SiO_{2}}+w_{TiO_{2}}d_{TiO_{2}}=m\protect\pi $ (black dotted lines), with $m=1,2$. (b) The $d_{SiO_{2}}$ and $d_{TiO_{2}}$ (blue dash dotted line) for which $\arg R_{per}=\arg r_{a1}$ is satisfied at $\kappa_{res}^{sym}$, when the losses are ignored, and the $d_{SiO_{2}}$ and $d_{TiO_{2}}$ (red solid line) for which $R_{per}=r_{a1}$ is satisfied at a $\kappa$ close to $\kappa_{res}^{sym}$, when the losses are considered. The large black dot indicates the choice of the parameters for the specific multilayer structure studied in the rest of this paper.}
\label{Fig:thicknesses}
\end{figure}

We consider addressing this task once a choice of multilayer materials has
been made, with $\varepsilon _{a}$ and $\varepsilon _{b}$ fixed. \ If we
then consider letting $d_{a}\rightarrow 0$, any existing band gap will
necessarily vanish, since the medium will become uniform, and at best we can
have $\left\vert A+D\right\vert =2$ in the limit. \ From \eqref{AplusD} we
see that as $d_{a}\rightarrow 0$ there are only discrete $d_{b}$ where this
will hold; only for these $d_{b}$ does the band gap survive until $d_{a}=0$.
\ We can plot these as points $(0,d_{b})$ in the plane of points $%
(d_{a},d_{b})$. \ Similarly, as $d_{b}\rightarrow 0$ there are only discrete 
$d_{a}$ for which $\left\vert A+D\right\vert =2$, signalling for which $%
d_{a} $ the band gap will survive until $d_{b}=0$. \ Connecting the
corresponding discrete points $(0,d_{b})$ and $(d_{a},0)$ in the $%
(d_{a},d_{b})$ plane by straight lines should then give a rough indication
of the location of the band gaps as $d_{a}$ and $d_{b}$ are varied. \ Those
straight lines are easily found to be identified by%
\begin{equation}
w_{a}d_{a}+w_{b}d_{b}=m\pi \hspace{0.5cm}\text{with}\hspace{0.5cm}%
m=1,2,3,.... \label{line_equation}
\end{equation}%
\ 

We illustrate this by considering periodic multilayers of SiO$_{2}$ and $%
TiO_{2}$, taking the dielectric constants of SiO$_{2}$ and TiO$_{2}$ as $%
\varepsilon _{SiO_{2}}=\varepsilon _{a}=2.0932$ \cite{Berini1} and $%
\varepsilon _{TiO_{2}}=\varepsilon _{b}=7.421$ \cite{refractive_index},
respectively. \ In Fig. \ref{Fig:thicknesses}a we plot the solutions of Eq. \eqref{line_equation} as dotted lines.
The solid lines indicate the solutions for the band edges ($\left\vert
A+D\right\vert =2$ for $d_{a}$ and $d_{b}$ in general both nonzero), with
the region in between each pair of lines indicating the values of $%
(d_{a},d_{b})$ for which there is a photonic bandgap. \ We see that the
dotted lines do indeed give a good indication of where in the $(d_{a},d_{b})$
plane the band gaps lie; we refer to the lines identified by Eq. \eqref%
{line_equation} as \textit{guide lines.} \ Note that the canonical
``quarter-wave stack'' with $w_{a}d_{a}=w_{b}d_{b}=\pi /2$, lies within the
first band gap and in fact is precisely on the line (\ref{line_equation})
with $m=1$.

To satisfy the matching condition \eqref{ra1} at $\kappa _{sym}^{res}$ we
must be in the band gap region ($\left\vert R_{per}\right\vert =\left\vert
r_{a1}\right\vert $) and have $\arg R_{per}=\arg r_{a1}$. \ It is easy to
determine where the latter condition is satisfied in the band gap region,
and we plot that as the blue dash dotted line in Fig. \ref{Fig:thicknesses}b, together with the solutions
of Eq. \eqref{line_equation}, again as dotted lines, where we focus on a region
of the $(d_{a},d_{b})$ plane where the $m=1$ guide line is close to the
center of the band gap region. \ Thus in the lossless
limit it is possible to choose a multilayer structure so that \eqref{ra1} is
exactly satisfied, and the LRSPs in the symmetric and antisymmetric
structures share the same $\kappa _{sym}^{res}$. \ We note that while the structures that do this are
characterized by values $(d_{a},d_{b})$ that lie
close to the guide lines, the solutions of $\arg R_{per}=\arg r_{a1}$ do not
run all the way to $d_{a}=0$ and $d_{b}=0$, as do the guide lines, for
before those limits are reached the solutions encounter the band edges.

We now reinstate loss in $\kappa _{sym}^{res}$ and in the water cladding. \
We no longer have $\left\vert R_{per}\right\vert =\left\vert
r_{a1}\right\vert $ automatically holding in the previously identified band
gap regions of the $(d_{a},d_{b})$ plane, as we did in the absence of loss,
and so to achieve \eqref{ra1} we would have to satisfy \textit{two}
nontrivial conditions, $\left\vert R_{per}\right\vert =\left\vert
r_{a1}\right\vert $ and $\arg R_{per}=\arg r_{a1}$. \ We can find curves in
the $(d_{a},d_{b})$ plane where each of these conditions is satisfied, but
for our choice of materials these curves do not intersect. \ So at least for
some choices of dielectric materials it is impossible to satisfy \eqref{ra1}
at $\kappa _{sym}^{res}$ in the ubiquitous presence of loss; we cannot
simply replace a uniform substrate with a periodic multilayer structure and
maintain the same LRSP.

Nonetheless, we can find complex values of $\kappa $ close to $\kappa
_{res}^{sym}$ where \eqref{ra1} \textit{is} satisfied. \ They can be
identified by choosing thicknesses $(d_{a},d_{b})$ close to one of the guide
lines, and searching in the complex plane for values of $\kappa $ that
satisfy \eqref{ra1}; we denote such solutions by $\kappa _{res}^{asym}$. \
In fact this is possible for a wide range of values $(d_{a},d_{b})$,
indicated by the red lines in Fig. \ref{Fig:thicknesses}b. \ The guide lines provide a good
indication of where the values $\left( d_{a},d_{b}\right) $ of interest
should be sought, although as in the lossless limit solutions $\kappa
_{res}^{asym}$ cannot be found all the way to $d_{a}=0$ and $d_{b}=0$. \
Note that in general the different points on the red line in Fig. \ref{Fig:thicknesses}b
correspond to different values of $\kappa _{res}^{asym}$, unlike the different points on the blue dash dotted line, which all correspond to solutions of Eq. \eqref{ra1} with the Fresnel
coefficients evaluated at $\text{Re}\kappa _{res}^{sym}$.

A reasonable design strategy is to adopt thicknesses $(d_{a},d_{b})$
associated with the center of the band gap region, resulting in an LRSP with
a field in the multilayer well-confined near the metal, and for which we can
expect a better tolerance for any fabrication errors. \ In line with this,
but still somewhat arbitrarily, we take $d_{SiO_{2}}=d_{a}=453.5nm$
and $d_{TiO_{2}}=d_{b}=161nm$ for the rest of this paper. \ This yields an $%
n_{res}^{asym}=\kappa _{res}^{asym}/\tilde{\omega}=$ $1.31824+i5.17\times10^{-5}$, corresponding to
a loss of 2.15dB/mm and an energy propagation length of about $2mm$. \ The
real part of $n_{res}^{asym}$ is very close to the real part of $%
n_{res}^{sym}$, and the loss for the asymmetric structure is actually
slightly less than for the LRSP of the original symmetric structure. \ Thus
while our original goal was to match $\kappa _{res}^{sym}$ and achieve the
low loss of a LRSP in a symmetric structure, we find that using a multilayer
structure it is possible to achieve even lower loss than in a symmetric
structure; we plan to return to this in future communications.

\begin{figure}[htbp]
\centerline{\includegraphics[width=6.5cm]{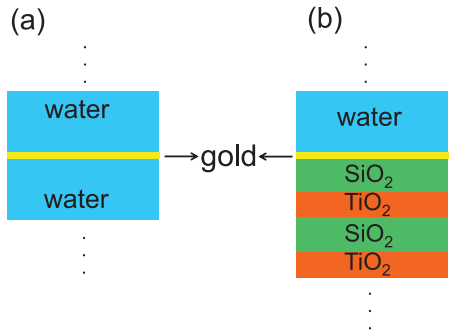}}
\caption{(a) An infinite symmetric structure supporting LRSP, with a $20nm$ gold layer. (b) An infinite asymmetric multilayer
structure supporting LRSP, with a $20nm$ gold layer, and a periodic multilayer structure of SiO$_{2}$ and TiO$_{2}$, with $d_{SiO_{2}}=453.5nm$
and $d_{TiO_{2}}=161nm$.}
\label{Fig:structure_infinite}
\end{figure}

\section{Symmetric and asymmetric LRSP fields}\label{LRSP_fields}

\begin{figure}[htbp]
\centerline{\includegraphics[width=7.5cm]{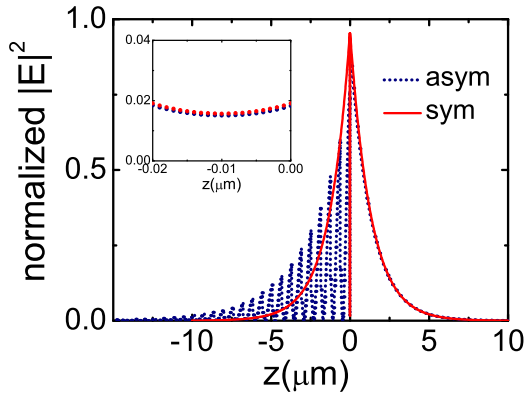}}
\caption{The normalized electric field intensity profile (see text below) of
the resonant mode of the asymmetric structure (dot blue line) and the LRSP
mode of the symmetric structure (solid red line). Inset: the zoomed
intensity profile in gold. The $z$ axis is normal to the plane of the
structure, and the (top) gold-water interface is at $z=0$.}
\label{Fig:fields}
\end{figure}

Some insight into the nature of the LRSPs in the infinite symmetric
and asymmetric structures we have considered (see Fig. \ref{Fig:structure_infinite}) can be gained by comparing the LRSP field profiles in the two structures. A fair comparison involves field profiles that are associated with the different structures and have the same field energy. Since loss is present, the field energy for a LRSP cannot be strictly defined, but since the loss is small in the sense that the LRSP
can propagate many wavelengths before decaying, we can proceed by replacing the complex quantity $1/\varepsilon(z,\omega)$, where $\varepsilon(z,\omega)$ is the position and frequency dependent dielectric constant, by Re$(1/\varepsilon(z,\omega))$, and using the standard expression for energy density in a dispersive medium to construct mode profiles corresponding to the same energy in both structures \cite{navin}. Scaling the electric field intensity (\textit{i.e.}, $|E|^{2}$) of the field profiles of the two structures in the same way yields the results shown in Fig. \ref{Fig:fields}.
The electric field intensity in the cladding ($z>0$) is evanescent, and
almost the same in the two structures. In the metal layer (inset), the
electric field intensity in the multilayer structure is similar to, and even
slightly smaller than, the field intensity in the symmetric structure; it is
for this reason that the loss of the mode in the asymmetric structure is
slightly smaller than that of the mode in the symmetric structure. While $%
|E|^{2}$ is symmetric about the center of the gold film in the symmetric
structure, it exhibits an evanescent envelope function in the multilayer in
the asymmetric structure. \ However, due to the multiple reflections and
interferences in the layers, there are oscillations in $|E|^{2}$ indicative of
the photonic band gap. Compared to an earlier proposed multilayer structure
for the LRSP \cite{Konopsky2}, in this structure the field is less confined,
but is also much smaller in the metal layer, and therefore the absorption
losses are smaller.

\begin{figure}[htbp]
\centerline{\includegraphics[width=6cm]{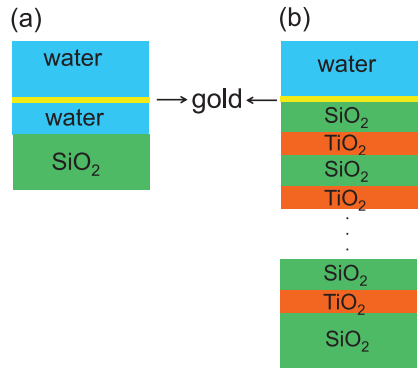}}
\caption{(a) A finite symmetric (b) finite asymmetric multilayer structure,
supporting LRSP.}
\label{Fig:structure_finite}
\end{figure}

In realistic structures the behaviour of the fields is modified by the
presence of substrates. \ While we continue to treat the cladding as
infinite, assuming that in sensing applications the thickness of the water
above the gold will be greater than the evanescent decay length of the field, we now consider the gold to be deposited on a finite multilayer structure with an SiO$_{2}$ substrate (see Fig. \ref{Fig:structure_finite}b). \
For comparison we also consider a structure with an SiO$_{2}$
substrate placed a finite distance below the gold film in an otherwise
symmetric structure (see Fig. \ref{Fig:structure_finite}a). \ The properties of the LRSP that
result in these structures can be studied by considering coupling into them
with an incident field from the substrate, which is the usual Kretschmann
configuration \cite{Kretschmann}. In such an excitation the field inside the
cladding is typically maximized when the rate of energy absorption in the
metal layer equals the rate of energy incident from the substrate, and thus
this usually identifies an optimum structure for sensing applications.
Although this design target can be identified by pole analysis strategies \cite{unpublished}, it can also be
found simply by examining a series of reflectivity calculations for different
thicknesses. For the symmetric structure we find that this critical coupling
occurs when the thickness of the water layer below the gold film is about $%
4.59\mu m$, and for the asymmetric structure we find that it occurs when the
number of periods is $25$, leading to a multilayer thickness of $15.36\mu m$%
. \ \ 

\begin{figure}[htbp]
\centerline{\includegraphics[width=7.5cm]{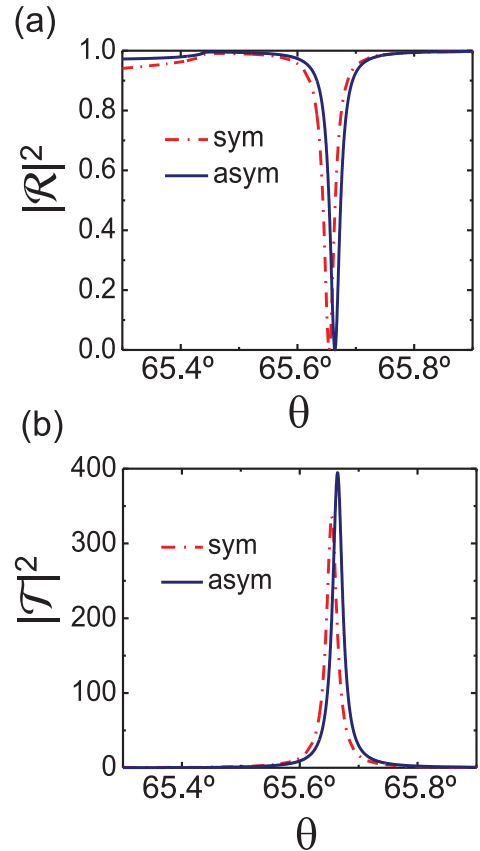}}
\caption{(a) Reflectivity $\left | \mathcal{R}\right |^2$, and (b) enhancement factor of the square of the field $\left | \mathcal{T} \right |^2$ (see text) for the symmetric (red dash dotted line) and the asymmetric (blue solid line) structures, both as a function of the angle of incidence from the substrate.}
\label{Fig:FC}
\end{figure}

In Fig. \ref{Fig:FC}a we plot the reflectivity $\left\vert \mathcal{R}%
\right\vert ^{2}$ of the two structures, each at its critical coupling
thickness, as a function of incident angle; $\mathcal{R}$ is the Fresnel
coefficient for $p$-polarized light incident from the substrate. The enhancement of the fields at the surface of each structure is determined by the Fresnel transmission coefficient from
the substrate to the cladding for $p$-polarized light, $\mathcal{T}$, and we
see from Fig. \ref{Fig:FC}b that enhancements in the square of the field of
the order of $400$ is expected. The narrow resonances shown in these plots indicate poles in the Fresnel coefficients $\mathcal{R}$ and $\mathcal{T}$. With respect to the poles in the corresponding infinite structures (see Fig. \ref{Fig:structure_infinite}), the real parts of the poles are shifted slightly and the imaginary parts increased (and thus the loss increased) due to radiative coupling into the substrate. A search in the complex plane finds the pole for the
symmetric structure at $\kappa =\tilde{\omega}n_{eff}^{sym}$, where $n_{eff}^{sym}=1.31814+i1.106\times10^{-4}$, with the imaginary part
corresponding to a loss of about 4.6dB/mm, or equivalently an energy
propagation length of about $0.94mm$, and the pole for the asymmetric
structure at $\kappa =\tilde{\omega}n_{eff}^{asym}$, where $%
n_{eff}^{asym}=1.31825+i1.053\times10^{-4}$, with the imaginary part
corresponding to a loss of about 4.39dB/mm, or equivalently an energy
propagation length of $1mm$. 

The dips in the reflectivities in Fig. \ref{Fig:FC}a occur at angles $\theta
^{sym,asym}$ associated with the real part of effective indices, $\text{Re}
n_{eff}^{sym}=n_{SiO_{2}}\sin \theta ^{sym}$ and $\text{Re}
n_{eff}^{asym}=n_{SiO_{2}}\sin \theta ^{asym}$ respectively, where $%
n_{SiO_{2}}=1.447$ is the index of refraction of the
substrate at $\lambda =1310nm$, and the widths of the dips are associated
with the imaginary parts of the effective indices. 
We note that the 
poles of the Fresnel coefficients extracted from equations such as Eq. \eqref{pole1} are used to approximate the values of Fresnel coefficients by pole
expansions valid for real $\kappa $, which are the appropriate $\kappa $ for
excitation in a Kretschmann configuration as discussed here, or for use in
superpositions to describe pulse propagation along a structure \cite{finite_beam}. \ Thus in
extracting these poles we should choose a definition of the square roots of $%
w_{i}(\kappa )$ such that the calculated values of the Fresnel coefficients
in the upper half of the complex $\kappa $ plane join continuously to those
calculated on the real $\kappa $ axis. \ Here this can be guaranteed by
choosing a branch cut that lies along the negative imaginary $\kappa $ axis.

\begin{figure}[htbp]
\centerline{\includegraphics[width=7.2cm]{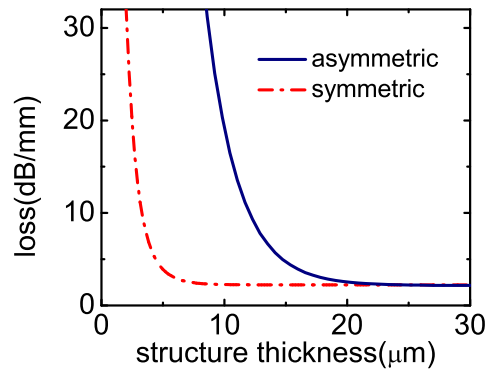}}
\caption{The loss of the LRSP in the symmetric (dash dot red line) and
asymmetric (solid blue line) structures, as the thickness of the structures
changes.}
\label{Fig:loss}
\end{figure}

If we turn from Kretschmann configurations to applications involving
end-fire coupling into LRSPs, optimization for sensing applications
typically involves the minimization of propagation loss. \ That loss
decreases as the thickness of the structure between the metal film and the
substrate increases, for as increasing thickness the radiative loss of
light into the substrate becomes less and less. In Fig. \ref{Fig:loss} the
calculated losses of the LRSPs in the symmetric and asymmetric structures
are shown as a function of the thickness of the structure, as extracted from
the poles of the Fresnel coefficients. In the symmetric structure, for a
water layer thickness larger than about $15\mu m$ the mode loss approaches a
limiting value of 2.23dB/mm, corresponding to the mode loss of the
infinite structure; in the asymmetric structure, for a multilayer thickness
larger than about $30\mu m$ ($49$ periods), the mode loss approaches a
limiting value of 2.15dB/mm, corresponding to the mode loss of the
infinite structure.

\section{Sensing with Planar Resonant Structures}\label{sensing}

\label{semianalytical} Resonant structures supporting guided modes, such as
those considered above, have their optical properties modified by the
presence of new species on or near the surface of the structure.\ This can
lead to their application as sensors, for the new species are located
precisely where the optical fields are largest, and thus their effect on the
properties of the guided modes can be significant. \ In this section we
derive a semi-analytic expression for a standard surface sensing parameter
that characterizes the effectiveness of such a sensor, and apply it to the
structures we have introduced in the last section.  \ 

\begin{figure}[htbp]
\centerline{\includegraphics[width=7.5cm]{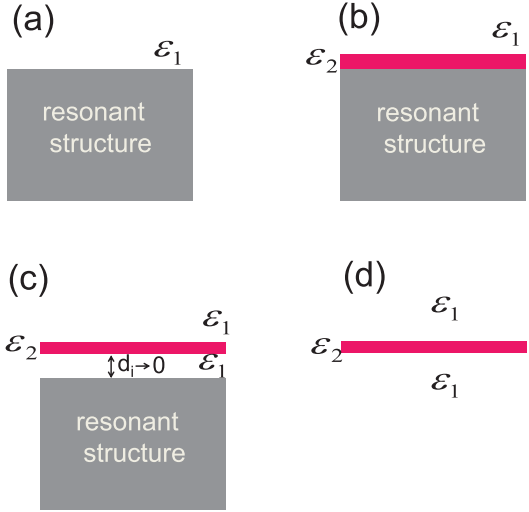}}
\caption{(a) A bare resonant structure, with a cladding of dielectric
constant $\protect\varepsilon _{1}$, (b) with a thin layer of molecules with
dielectric constant $\protect\varepsilon _{2}$ on top, (c) an
infinitesimally thin layer of medium 1 under the molecular layer, and (d)
the molecular layer between to medium with dielectric constant $\protect%
\varepsilon _{1}.$}
\label{Fig:resonant}
\end{figure}

We begin generally and consider an arbitrary planar resonant structure (Fig. 
\ref{Fig:resonant}a), supporting a resonant mode at a complex wavenumber $%
\kappa _{res}^{0}=\kappa _{R}+i\kappa _{I}$. If a thin molecular layer with
effective dielectric constant $\varepsilon _{2}$ is placed on the structure
(Fig. \ref{Fig:resonant}b), the complex wavenumber of the mode shifts, and a
surface sensing parameter can be defined \cite{berini_sensing} as 
\begin{equation}
G=\frac{1}{\kappa _{I}}\frac{\partial }{\partial d}\text{Re}(\Delta \kappa
_{res}),  \label{sensing_parameter}
\end{equation}%
where $\Delta \kappa _{res}=\kappa _{res}^{m}-\kappa _{res}^{0}$, and $%
\kappa _{res}^{m}$ is the complex wavenumber of the mode in the presence of
the molecular layer. 
For $\kappa $ close to $\kappa _{res}^{0}$ we can use a pole expansion \cite%
{finite_beam,aida_spont} for the reflection coefficient from the cladding to
the bare resonant structure, which we now denote very generally by $\bar{R}%
_{1s}$, such that 
\begin{equation}
\bar{R}_{1s}\simeq \frac{\rho _{1s}}{\kappa -\kappa _{res}^{0}},
\label{pole_expansion}
\end{equation}%
where $\rho _{1s}$ characterizes the pole strength and is in general
complex. For the resonant structure with the molecular layer on top,
modelled as a thin dielectric film, we can construct an expression for the
Fresnel reflection coefficient $\bar{R^{\prime }}_{1s}$ in terms of $\bar{R}%
_{1s}$ by adding an infinitesimally thin layer, with a thickness $%
d_{i}\rightarrow 0$ and a dielectric constant $\varepsilon _{1}$, just below
the molecular layer (Fig. \ref{Fig:resonant}c). Then 
\begin{equation}
\bar{R^{\prime }}_{1s}=R+\frac{T\bar{R}_{1s}T}{1-R\bar{R}_{1s}},
\label{help1}
\end{equation}%
where $R$ and $T$ are the Fresnel reflection and transmission coefficients
for the molecular layer sandwiched between two media with dielectric constant 
$\varepsilon _{1}$ (Fig. \ref{Fig:resonant} d). Inserting Eq. %
\eqref{pole_expansion} in Eq. \eqref{help1}, 
\begin{equation}
\bar{R^{\prime }}_{1s}=R+\frac{T\rho _{1s}T}{\kappa -(\kappa
_{res}^{0}+R\rho _{1s})},
\end{equation}%
implying that the complex wavenumber of the mode in the presence of
molecules, $\kappa _{res}^{m}$, is $\kappa _{res}^{m}=\kappa
_{res}^{0}+R\rho _{1s}$, and the shift in the complex wavenumber is 
\begin{equation}
\Delta \kappa _{res}=R\rho _{1s}.  \label{shift}
\end{equation}%
The pole strength $\rho _{1s}$ is a parameter of the bare resonant
structure, and does not depend on the properties of the molecular layer; in
general it must be determined numerically. The reflection coefficient $R$
(see Fig. \ref{Fig:resonant}d), however, is 
\begin{equation}
R=r_{12}+\frac{t_{12}r_{21}t_{21}e^{2iw_{2}d}}{1-r_{21}r_{21}e^{2iw_{2}d}},
\label{help2}
\end{equation}%
where $d$ is the thickness of the molecular layer, and $r_{12}$, $t_{12}$, $%
r_{21}$, and $t_{21}$ are the Fresnel reflection and transmission
coefficients between the cladding and the molecular layer (recall \eqref{spol}%
,\eqref{ppol}). Inserting Eq. \eqref{spol} or \eqref{ppol}
into Eq. \eqref{help2}, for thin molecular layers, where 
\begin{equation}
(w_{1}\pm w_{2})d\ll 1,  \label{approximation}
\end{equation}%
for $s$-polarization we find 
\begin{equation}
R\simeq \frac{n_{os}}{1-n_{os}},  \label{help3}
\end{equation}%
with 
\begin{equation*}
n_{os}=\frac{i\tilde{\omega}^{2}}{2w_{1}}(\varepsilon _{2}-\varepsilon
_{1})d,
\end{equation*}%
while for $p$-polarization 
\begin{equation}
R\simeq \frac{n_{-}}{1-n_{+}},  \label{help4}
\end{equation}%
where 
\begin{equation*}
n_{\pm }=\frac{i\kappa ^{2}}{2w_{1}}\frac{(\varepsilon _{2}-\varepsilon
_{1})d}{\varepsilon _{2}}\pm \frac{iw_{1}}{2}\frac{(\varepsilon
_{2}-\varepsilon _{1})d}{\varepsilon _{1}}.
\end{equation*}%
These calculations agree with the similar calculations presented earlier by
Cheng et al. \cite{rangan}; using Eq. \eqref{help3} or \eqref{help4}, and %
\eqref{shift} in Eq. \eqref{sensing_parameter}, we find a semi-analytic
expression for the surface sensing parameter in terms of the thickness of
the molecular layer, the dielectric constants of the molecular layer and the
cladding, and the pole strength. We assume a molecular layer with an
effective index of refraction of $1.5$ and thickness up to $10nm$, and
calculate the sensing parameters for the symmetric and asymmetric structures
discussed in the previous section, both exactly by numerically determining
the shift $\Delta \kappa _{res}$ in the position of the pole of the full
structures, and approximately from the semi-analytic expressions presented
in this section; here the $p$-polarized expressions are the relevant ones. 
\begin{figure}[htbp]
\centerline{\includegraphics[width=7.5cm]{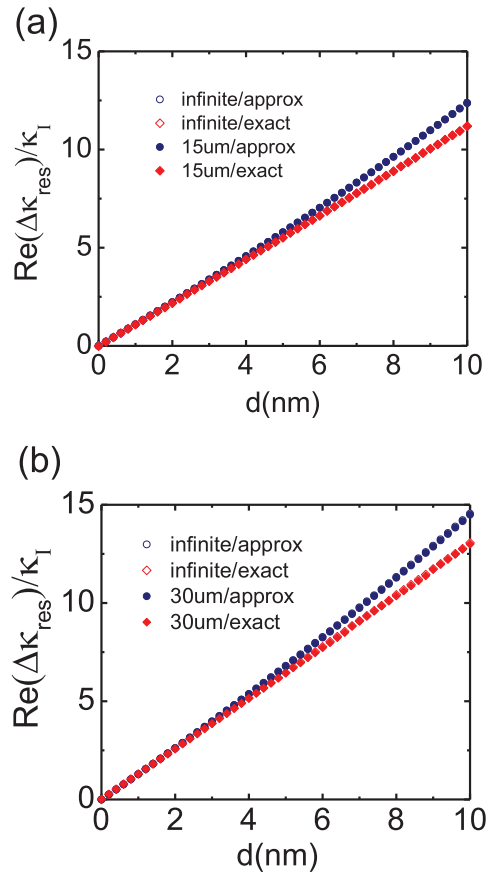}}
\caption{The shift in the resonance wavenumber divided by the width of the
resonance for the (a) infinite and $15\protect\mu m$ symmetric and (b)
infinite and $30\protect\mu m$ asymmetric structures. The circles and the
diamonds are the semi-analytic and exact calculations, respectively. The
hollow circles and diamonds correspond to the infinite structures
calculations, and overlap with the full circles and diamonds that correspond
to the finite structures calculations.}
\label{Fig:compare}
\end{figure}
In Fig. \ref{Fig:compare} we show the results for $\text{Re}(\Delta \kappa
_{res})/\kappa _{I}$ of the infinite symmetric and asymmetric structures.
The semi-analytic calculations match with the exact calculations for $%
d<5nm$, but as the thickness of the molecular layer increases the assumption
(\ref{approximation}) loses its validity and the approximated calculations
deviate more from the exact calculations; nonetheless, they remain accurate
to about $10\%$ for thicknesses as large as $10nm$. \ The slope of the
curves in Fig. \ref{Fig:compare} around $d=0$ gives $G$, and we find $%
G=1.09nm^{-1}$ and $G=1.28nm^{-1}$ for the infinite symmetric and asymmetric
structure, respectively; the larger $G$ for the asymmetric structure is due
to the slightly smaller resonance width ($\kappa _{I}$) in the asymmetric
structure. From Fig. \ref{Fig:loss}, it is clear that if the thickness of
the finite symmetric (or asymmetric) structure is larger than about $15\mu m$
(or $30\mu m$), the loss in the structure is about the loss in the
corresponding infinite structure. 
As expected, the results for $\text{Re}(\Delta \kappa _{res})/\kappa _{I}$ for a $15\mu m$ symmetric structure and a $30\mu m$ asymmetric structure are indistinguishable from the corresponding infinite structures, as shown in  Fig. \ref{Fig:compare}.
\begin{figure}[htbp]
\centerline{\includegraphics[width=7.5cm]{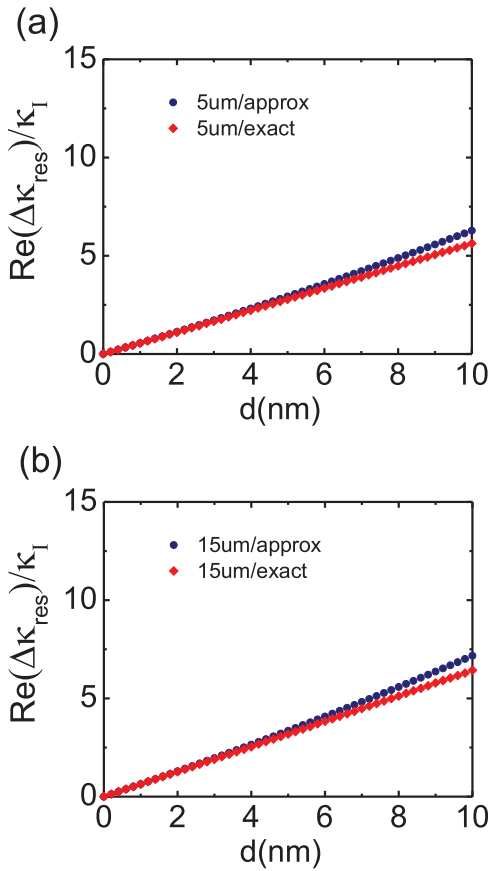}}
\caption{The shift in the resonance wavenumber divided by the width of the
resonance for the (a) $4.590\protect\mu m$ symmetric and (b) $15.362\protect%
\mu m$ asymmetric structures. The circles and the diamonds are the
semi-analytic and exact calculations, respectively.}
\label{Fig:critical_compare}
\end{figure}
We have done similar calculations for the finite symmetric and asymmetric
structures at thicknesses corresponding to the critical coupling thickness ($%
4.590\mu m$ for the symmetric structure and $15.362\mu m$ for the asymmetric
structure), and the result is shown in Fig. \ref{Fig:critical_compare}.
These would be relevant for sensing in a Kretschmann configuration rather
than an end-fire coupling configuration. We find $G=0.55nm^{-1}$ and $%
G=0.63nm^{-1}$ for the finite symmetric and asymmetric structures,
respectively, which are smaller than the corresponding values shown in
Fig. \ref{Fig:critical_compare}, as the coupling losses are larger here.

\section{Conclusion}

In this paper we have presented a new strategy for designing asymmetric
multilayer structures that support LRSPs. We have shown that if the Fresnel
reflection coefficient from the cladding to the multilayer structure below
the metal film vanishes at the complex wavenumber of the LRSP in a symmetric
structure, the LRSP resonance condition for the symmetric and asymmetric
multilayer structures become the same, and the asymmetric structure supports
a LRSP equivalent to that supported in the symmetric structure. \ For an
arbitrary choice of materials for the multilayer structure it is impossible
to satisfy this condition exactly, but at complex wavenumbers close to that
of the LRSP in the symmetric structure a resonance condition can be found,
in some instances with even less loss than that of the LRSP in the symmetric
structure. \ 

We have provided a protocol for determining this resonance condition based
on first describing a model system without loss, and then including the loss
in the final design. We have also studied how the losses depend on the
thickness of the multilayer structure, taking into account radiative
contributions to the substrate. For a multilayer of SiO$_{2}$ and 
TiO$_{2}$ we have found that the radiative losses are negligible if the
multilayer structure thickness is about or greater than $30\mu m$. For
biosensing applications involving an arbitrary planar resonant structure, we
have derived a semi-analytic expression for a standard surface sensing
parameter identifying the dependence of the sensing parameter on the
dielectric constant of the molecular layer, its thickness, and the original
pole strength of the resonance on the bare structure; for typical parameters
we find that there is a good match between these semi-analytic expressions
and the exact results if the thickness of the molecular layer is less than
about $5nm$, with corrections only on the order of about $10\%$ for
molecular layer thicknesses up to $10nm$. \ For a $20nm$ gold film the
surface sensing parameter for the $30\mu m$ thick multilayer structure is $%
G=1.28nm^{-1}$, larger than the value of $G=1.09nm^{-1}$ for a $15\mu m$
thick symmetric structure.

Compared to the $20nm$ gold structure of Min et al. \cite{Berini1}, which
has an intensity attenuation of 3.26dB/mm and a surface sensing parameter $%
G=1.29nm^{-1}$, our asymmetric multilayer structure has a smaller loss
(minimum of 2.15dB/mm) and a similar sensing parameter ($G=1.28nm^{-1}$).
However, the structure of Min et al. \cite{Berini1} is a thin film suspended
in air, and compared to our multilayer structure is expected to be less
stable and harder to fabricate. Compared to the multilayer structure
studied by Konopsky et al. \cite{Konopsky2}, the multilayer structure we
present here is fully periodic, and does not require an additional layer
between the metal and the periodic multilayer. However, a direct comparison
of the losses and the surface sensing parameters of these two structures is
not possible, as the metal layer in the work of Konopsky et al. \cite%
{Konopsky2} is palladium, which is lossier than gold. Nevertheless, if the
number of the periods in that multilayer structure is increased from $14$
periods, the mode losses can be decreased by a factor of two. \ More
generally, the design strategy presented here can be applied to a range of
structures involving other metals and other multilayers to systematically
explore the parameter space and optimize the predicted behavior.

In previous work \cite{aida_spont}, we calculated Raman scattering from
molecules on planar resonant dielectric structures, and showed that the Raman
signal is enhanced when the pump field couples to a resonant mode. The
asymmetric multilayer structure we studied in this paper can also be used as
a substrate for surface enhanced Raman scattering (SERS) \cite{sers}, when the
pump field is coupled to the LRSP excitation. In particular, we expect that
the good surface functionalization of gold films may make these structures
more promising SERS substrates than fully dielectric multilayer structures 
\cite{srini}.

\section{Acknowledgements}
This work was supported by BiopSys: the Natural Sciences and Engineering Research Council of Canada Strategic Network for Bioplasmonic Systems.

\end{document}